\documentclass[9pt,twocolumn]{article}
\usepackage{times,bm}
\usepackage[affil-it,]{authblk}
\usepackage[pagebackref=false,colorlinks=true,linkcolor=red,citecolor=red,urlcolor=black]{hyperref}

\usepackage{geometry}
\usepackage[switch*, displaymath,pagewise, mathlines]{lineno}
\usepackage{graphicx}
\usepackage{cite}
\usepackage{amsmath}
\usepackage{amsfonts}
\usepackage{amssymb}
\usepackage{slashed}
\usepackage{subfloat}
\usepackage{slashed}
\usepackage{lipsum}
\usepackage{color}
\usepackage{float}
\usepackage{multirow,multicol}
\usepackage{tabulary}
\usepackage[flushleft]{threeparttable}
\usepackage{pgfplots,subfigure}
\usepackage{xcolor}
\numberwithin{equation}{section}
\usepackage{tikz}
\usepackage{ulem}
\usepackage{hyperref}
\usepackage{nameref}
\usepackage{comment}

\usepgflibrary{arrows}
\usetikzlibrary{shapes.callouts}
\tikzset{
	level/.style   = { thick, },
	connect/.style = { dotted, red   },
	notice/.style  = { draw, rectangle callout, callout relative pointer={#1} },
	label/.style   = { text width=2cm }
}

\usepackage{letltxmacro}
\makeatletter
\let\oldr@@t\r@@t
\def\r@@t#1#2{%
	\setbox0=\hbox{$\oldr@@t#1{#2\,}$}\dimen0=\ht0
	\advance\dimen0-0.2\ht0
	\setbox2=\hbox{\vrule height\ht0 depth -\dimen0}%
	{\box0\lower0.4pt\box2}}
\LetLtxMacro{\oldsqrt}{\sqrt}
\renewcommand*{\sqrt}[2][\ ]{\oldsqrt[#1]{#2}}
\makeatother
\usepackage{environ}
\begin{document}

\newcommand{{\ri}}{{\rm{i}}}
\newcommand{{\Psibar}}{{\bar{\Psi}}}

\title{Effect of internal magnetic flux on a relativistic spin-1 oscillator in the spinning point source-generated spacetime}
\author{ \textit {Abdullah Guvendi}$^{\ 1}$\footnote{\textit{E-mail: abdullah.guvendi@erzurum.edu.tr}}~,~ \textit {Semra Gurtas Dogan}$^{\ 2}$\footnote{\textit{E-mail: semragurtasdogan@hakkari.edu.tr} (Corresponding author)}  \\
	\small \textit {$^{\ 1}$Department of Basic Sciences, Faculty of Science, Erzurum Technical University, 25050, Erzurum,
		Turkey}\\
	\small \textit {$^{\ 2}$Department of Medical Imaging Techniques, Hakkari University, 30000, Hakkari, Turkey}\\}

\date{}
\maketitle

\begin{abstract}
We \textcolor{black}{consider a} charged relativistic spin-1 oscillator under the influence of an internal magnetic flux in a 2+1 dimensional spacetime induced by a spinning point source. In order to \textcolor{black}{analyze} the effects of the internal magnetic flux and spin of the point source on the relativistic dynamics of such a vector field, we seek a non-perturbative solution of the associated spin-1 equation derived as an excited state of Zitterbewegung. By performing an analytical solution of the resulting equation, we determine exact results for the system in question. Accordingly, we \textcolor{black}{analyze} the effects of spin of the point source and internal magnetic flux on the relativistic dynamics of the considered test field. \textcolor{black}{We see that the spin of such a field can be altered by the magnetic flux and this means that the considered system may behave as a fermion or boson according to the varying values of the magnetic flux, in principle}. We observe that the internal magnetic flux and the spin of the point source impact on the relativistic energy levels and probability density functions. \textcolor{black}{Also, our results indicate} that the spin of the point source breaks the symmetry of the energy levels corresponding to particle-antiparticle states.
\end{abstract}

\begin{small}
\begin{center}
\end{center}
\end{small}


\bigskip

\section{Introduction}\label{sec1}

\textcolor{black}{Analyzing the effects of curved spaces on the relativistic dynamics of the quantum mechanical systems is of great importance because such investigations may serve to build a complete theory combining the fundamental areas such as the quantum ($\hbar$), relativity ($c$) and gravity ($\mathcal{G}$) theories of modern physics. In the mathematical framework, \textcolor{black}{one of the useful ways} to determine the effects of spacetime topology and gravity on the relativistic quantum mechanical systems is to use the relativistic wave equations such as the Dirac equation, Duffin-Kemmer-Petiau ($\mathcal{DKP}$) equation, Klein-Gordon equation and fully-covariant many-body equations \cite{22,n-6,n-7,n-8,n-9,n-10,n-2,n-3,n-4,n-5,n-11}. Especially for spinning particles, these equations are different possible quantum states of the same classical system because they can be derived through the canonical quantization of the action for classical spinning particle with Zitterbewegung \cite{1}. One can see that the covariant spin-1 equation can be derived as an excited state of Zitterbewegung by applying the same quantization procedure provided that the spinor is a symmetric spinor of rank-two \cite{1}. In its generalized form, this equation possesses spin algebra constructed through Kronecker product of the generalized Dirac matrices with the unit matrix and accordingly it is quite different from the Dirac equation. Briefly, the corresponding spin-1 field can be constructed by the direct product of two symmetric Dirac fields but the Dirac equation and the covariant spin-1 equation are equivalent to each other in  the $1+1$ dimensions (except for the mass factors). This is because we get rid of the spin effects in $1+1$ dimensions. However, the aforementioned spin-1 equation corresponds to the} spin-1 sector of the well-known $\mathcal{DKP}$ equation in $2+1$ dimensions \cite{1,2,3,4}. \textcolor{black}{That is, the spin-1 equation provides serious mathematical convenience for analyzing the dynamics of the corresponding quantum mechanical systems possessing natural dynamical symmetry}. It is known that one of the exactly soluble systems in the non-relativistic quantum mechanics is the harmonic oscillator and this system is described by adding a quadratic interaction potential into the free Schrodinger Hamiltonian \cite{book1}. However, generalization of a quantum oscillator is not unique in relativistic quantum theory \cite{5}. The Dirac oscillator ($\mathcal{DO}$) was described by modifying the momentum operator(s) \cite{DO} of the free Dirac Hamiltonian and it has been shown that this non-minimal interaction has important potential for many applications \cite{6,7,8,9,10,11,12}. At first look, one can see that this modified Dirac equation is linear in both momentum and coordinate and its solutions yield an ordinary quantum oscillator plus a very strong spin-orbit coupling in the non-relativistic limit \cite{DO}. Hence, it is called as $\mathcal{DO}$. \textcolor{black}{By determining the electromagnetic potential corresponding to the $\mathcal{DO}$ interaction it was shown that the $\mathcal{DO}$ corresponds to a real physical system describing the interaction of the anomalous magnetic moment with a linearly growing electric field} \cite{7}. Due to this result, the $\mathcal{DO}$ system was regarded as an alternative model for the quarks in Quantum Chromodynamics \cite{7,8,9,10,11,12,13,14}. Through some analogies, other relativistic oscillators have introduced and studied in many areas of modern physics \cite{12,15,16,17,n-2,n-3,n-4,n-5}. For instance, such systems are used to estimate mass spectra for mesons and baryons \cite{10,11,12,13,14}. \textcolor{black}{Furthermore, relativistic oscillators are excellent mathematical tools  for determining} the effects of the background on the corresponding physical systems \cite{18,19,20,21} because they are exactly soluble systems (in general) besides the low-energy bound state systems in Quantum Electrodynamics \cite{22,22-b} (see also \cite{23}). Recently, \textcolor{black}{by taking into account the spin algebra}, the relativistic spin-1 oscillator has been introduced by adding a non-minimal interaction term into the mentioned relativistic spin-1 equation and the results of a number of applications were announced \cite{24,25,26,27,28}. \textcolor{black}{Moreover, if one gets rid of the spin effects, it can be seen that the announced results are exactly the same (except for the mass factors) with the previously obtained results for a 1+1-dimensional $\mathcal{DO}$ system \cite{pacheco} and for the singlet quantum state of a fermion-antifermion pair interacting through the $\mathcal{DO}$ coupling \cite{10}}. The other applications of the relativistic spin-1 equation can be found in the Refs. \cite{29,30,31,32}. However, we could not find any published results based on the interaction of a charged relativistic spin-1 oscillator with the electromagnetic fields in the current literature. Hence, in this contribution, \textcolor{black}{our purpose is to analyze} the effect of internal magnetic flux on a relativistic spin-1 oscillator in the $2+1$ dimensional background geometry spanned by a spinning point source \cite{34}. \textcolor{black}{This also allows us} to discuss the influence of the background on such a vector test field. In $2+1$ dimensions, the solutions of Einstein equations for static one-body and many-body sources can be obtained by applying geometrical and analytical methods \cite{33,34,35,36}. In the one-body case, it was shown that the surface metrically lies outside the origin \cite{20,33,34,35,36}. \textcolor{black}{Stationary spinning source and static point source generated spacetime backgrounds were first introduced in $2+1$ dimensions through the mentioned solutions \cite{33,34,35,36} and then these solutions were extended to $3+1$ dimensions where there is a dynamical symmetry \cite{33,34,35,36}. That is, dynamics of a vector field remains invariant under the Lorentz boost on the added third spatial coordinate ($z$-axis in the cylindrical coordinates) \cite{18}}. In $2+1$ dimensions, the source locates at the spatial origin (intersection point of the string with the space-like subspace) and $2+1$ dimensional cosmic string-generated geometric backgrounds are known also as point source-induced spacetime backgrounds \cite{33,34,35,36}. These objects are responsible for some non-trivial topologies because such an induced spacetime is not flat globally even though it is flat locally. Accordingly, the cosmic strings or point sources change the dynamics of the quantum systems by altering the symmetry of the spacetime. \textcolor{black}{Such a spacetime background can be characterized by two parameters. First of them is the angular deficit parameter ($\alpha \in (0, 1]$) that relates with the tension of the source and the other is the spin parameter ($\varpi$) of the source. Furthermore, determining the effects of such a non-trivial topology on the relativistic dynamics of the physical systems are very important in the context of cosmology and particle physics. In addition, a point source-generated background can be useful to discuss the dynamics of particles in a graphene-like structures described by a surface of constant negative curvature \cite{20}. Here, we should note that such a point source-generated spacetime describes an anti-conical spacetime if $\alpha>1$. This also allows us to discuss some conceptual issues. In short, by considering $\alpha=1$ and $\varpi=0$, any obtained result for a physical system in such a point source-generated spacetime gives the result for the same physical system in a globally and locally flat spacetime without losing any information.}

On the other hand, in electromagnetic theories it is known that a charged ($e$) particle moving between the determined two points within a certain path in a region without magnetic field experiences a phase shift \cite{19}. This means that such a particle will acquire a different phase when the path is changed. Such an effect can occur when the particle feels an internal magnetic flux and hence the corresponding wave function may acquire a phase shift in this scenario. This effect, also known as the Aharonov-Bohm effect, can be observed by considering the interaction of a charged particle with  internal magnetic flux \cite{19}. This is a non-local effect that can be investigated through the gauge-dependent vector potential \cite{19}. The effects of the internal magnetic flux on the relativistic quantum mechanical systems have been widely studied. Dynamics of the $\mathcal{DO}$ in the presence of an internal magnetic flux in a $2+1$-dimensional Gödel-type background spacetime \cite{19}, linear confinement of a spin-0 oscillator interacting with a uniform magnetic field and Aharonov–Bohm potential \cite{37}, the effect of internal magnetic flux on a generalized $\mathcal{DO}$ in the cosmic dislocation background geometry \cite{38}, the interaction of a  generalized $\mathcal{DKP}$ oscillator with an internal magnetic flux in the Som–Raychaudhuri background spacetime \cite{39}, and the Aharonov-Bohm effect on a charged scalar field in the screw dislocation background \cite{40}(see also \cite{n-1}) can be considered among such investigations. Here, it is worth mentioning that the spin-0 sector of the $\mathcal{DKP}$ equation was studied to analyse the effect of the internal magnetic flux on the associated relativistic oscillators. \textcolor{black}{In other words, the number of studies based on the dynamics of the relativistic spin-1 oscillator is relatively few. Therefore, determining the effects of an internal magnetic flux on a charged relativistic spin-1 oscillator in a spinning point source-induced spacetime can be very useful for discussing some conceptual issues besides its possible applications, in principle. As aforementioned before, we can analyze the relativistic dynamics of the considered system by obtaining an analytical solution of the associated covariant spin-1 equation since the considered system has a dynamical symmetry (see \cite{18}).}

This paper is structured as follows: in sec. (\ref{sec2}) we introduce the corresponding form of the covariant spin-1 equation for a charged spin-1 oscillator under the effect of an internal magnetic flux in a 2+1 dimensional spinning point source-induced spacetime. In sec. (\ref{sec3}), we obtain a set of coupled equations and arrive at a $2^{nd}$ order wave equation for the considered test field. Then, we perform an analytical solution of the resulting wave equation and analyze the effects of the internal magnetic flux as well as the spin of point source on the dynamics of the system under scrutiny. In sec. (\ref{Conc}), we give a summary and discuss the results in \textcolor{black}{detail}.

\section{Mathematical procedure} \label{sec2}
\begin{figure*}
\centering
\subfigure[$\varpi=1$]{\includegraphics[scale=0.80]{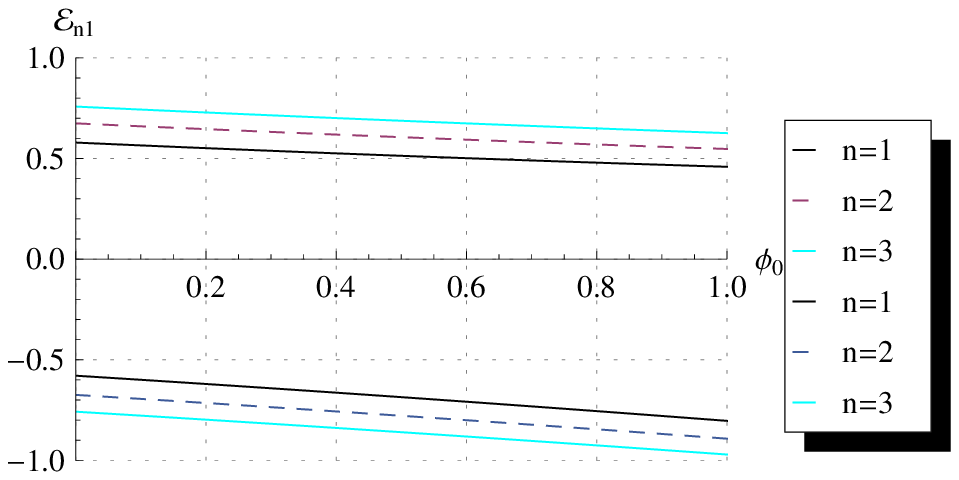}}\quad \subfigure[$\phi_{0}=0.5$]{\includegraphics[scale=0.80]{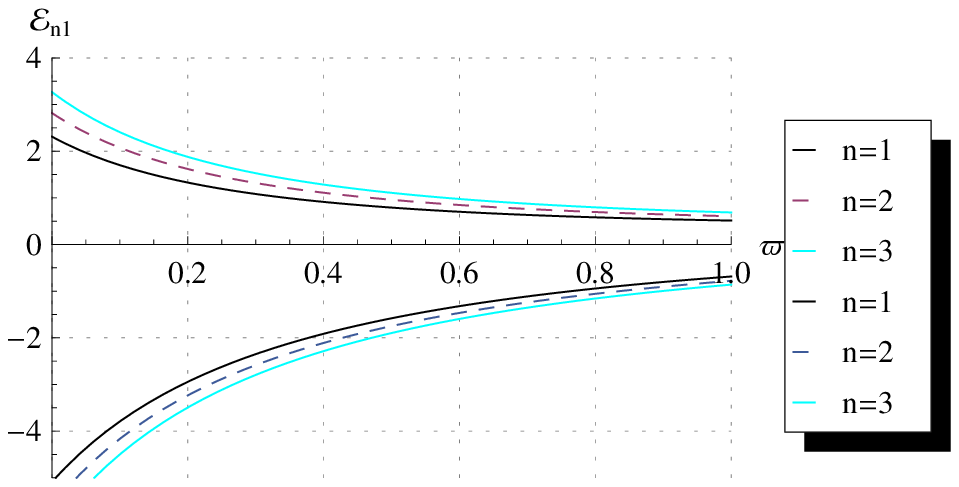}}\\
\caption{Effects of the internal magnetic flux and spin parameter of the point source on the relativistic energy levels. Here $e=1$, $c=1$, $\hbar=1$, $m=1$, $\alpha=0.9$, $\omega_{o}=1$, $\pi=1$. }
\label{figa}
\end{figure*}

In this section, we introduce the generalized relativistic spin-1 equation and try to derive a set of coupled equations for a charged relativistic spin-1 oscillator exposed to an internal magnetic flux in a spinning point source-generated $2+1$-dimensional spacetime background. \textcolor{black}{In 2+1 dimensions, the generalized form of the spin-1 equation can be written as the following \cite{1,2,3,4}}
\begin{eqnarray}
&\left\lbrace \beta^{\mu}\slashed{\mathcal{D}}_{\mu}+i\tilde{m}\mathcal{I}_4  \right\rbrace \Psi\left(\textbf{x}\right) =0,\nonumber\\
&\beta^{\mu}=\frac{1}{2}\left(\gamma^\mu \otimes \mathcal{I}_2+\mathcal{I}_2 \otimes \gamma^\mu \right),\quad \tilde{m}=\frac{mc}{\hbar},\nonumber\\
&\slashed{\mathcal{D}}_{\mu} =\left[ \partial_\mu +i\frac{eA_\mu}{\hbar c}-\Omega_{\mu} \right], \Omega_{\mu}= \Gamma_\mu \otimes \mathcal{I}_2+\mathcal{I}_2\otimes \Gamma_\mu,\nonumber\\
& \label{vbe}
\end{eqnarray}
in which \textcolor{black}{the Greek index $\mu$ refers to the coordinates of the curved spacetime}, $\gamma^{\mu}$ are the space-dependent Dirac matrices, $\Gamma_{\mu}$ are the spinorial affine connections for spin-$1/2$ field, $\Psi(\textbf{x})$ is the spin-$1$ field, with mass of $m$, \textcolor{black}{constructed through Kronecker product of symmetric two Dirac fields \cite{1}}, $\hbar$ is the reduced Planck constant, $c$ is the speed of light, $\textbf{x}$ is the spacetime position vector, the objects $\mathcal{I}_{2}$ and $\mathcal{I}_{4}$ stand for the $2\times2$ and $4\times4$-dimensional identity matrices, respectively, and the symbols $\otimes$ are used to indicate the Kronecker (direct) products. \textcolor{black}{Now, to derive the corresponding form of the fully-covariant spin-1 equation for a charged spin-1 oscillator in the 2+1 dimensional spinning cosmic string spacetime, let we start with by introducing the 2+1 dimensional spacetime generated by a spinning point source. This background can be represented through the following spacetime interval \cite{22,33,34,35,36}}
\begin{eqnarray}
&ds^{2}=c^2dt^{2}+2c\varpi dt d\phi-dr^{2}-(\alpha^{2}r^{2}-\varpi^2)d\phi^{2},\nonumber\\
&\alpha=\left( 1-\frac{4G\mu_{s}}{c^2}\right),\quad \varpi=\frac{4j_{s}G}{c^3}.\label{metric}
\end{eqnarray}
Here, \textcolor{black}{the parameter $\alpha$, depending on the mass density ($\mu_{s}$) of the string, relates with the angular deficit in the background and the letter $G$ stands for the well-known Newtonian gravitational constant. Provided $0<\frac{4G\mu_{s}}{c^2}<1$, the spacetime background is not flat when viewed globally although it is flat locally. As we mentioned before, the corresponding surface becomes conical if $0<\alpha<1$ and it becomes anti-conical when $\alpha>1$ \cite{33,34,35,36}. Also, the $\varpi$ refers to the spin of the point source and depends on the angular momentum density ($j_s$) of the cosmic string. It is clear that the Eq. (\ref{metric}) describes a flat Minkowski spacetime in terms of polar coordinates with the signature $(+,-,-)$ if $\mu_{s}=0$ and $j_{s}=0$. This means that the mass density and spin of the cosmic string source alter the topology and accordingly it changes the symmetry of the background where the angle $\phi$ runs in the interval $0\leq\phi<2\pi$. However, it should be said that the $\alpha>1$ does not make any sense in the context of general relativity. As is mentioned, dynamics of a physical system in a cosmic string spacetime can be investigated in 2+1 dimensions without loss of generality because the dynamics of the considered system remains invariant under the Lorentz boost on the added third spatial coordinate ($z$ axis in the cylindrical coordinates)}. The spinor field must be described locally by using the rule of local Lorentz transformations. Thus, we need to find a local reference frame according to the observers. We can construct the reference frame through a non-coordinate basis $\hat{\theta}^{a}=e^{a}_{\,\,\,\mu}(x)dx^{\mu}$, in which the components  $e^{a}_{\,\,\,\mu}(x)$ are the vierbeins and satisfy the following condition $g_{\mu\nu}(x)=e^{a}_{\,\,\,\mu}(x)e^{b}_{\,\,\,\nu}(x)\eta_{ab}$ where the $\eta_{ab}=\mathrm{diag}(+,-,-)$ represents the flat Minkowski metric. Also, by using the relation: $dx^{\mu}=e^{\mu}_{\,\,\,a}(x)\hat{\theta}^{a}$ we can find the inverse of vierbeins. Here, we should note that the vierbeins and their inverses obey the following two conditions: $e^{\mu}_{\,\,\,a}(x)e^{a}_{\,\,\,\nu}(x)=\delta^{\mu}_{\,\,\,\nu}$ and $e^{a}_{\,\,\,\mu}(x)e^{\mu}_{\,\,\,b}(x)=\delta^{a}_{\,\,\,b}$ where $\mu,\nu=t,r,\phi$ and $a,b=0,1,2$. By this way, one can obtain the inverse of vierbeins as (see also \cite{22})
\begin{eqnarray}
&e^{t}_{\,\,\,0}(x)=\frac{1}{c},\quad e^{r}_{\,\,\,1}(x)=1,\quad e^{t}_{\,\,\,2}(x)=-\frac{\varpi}{c\alpha r}, \nonumber\\
&e^{\phi}_{\,\,\,2}(x)=\frac{1}{\alpha r}.\label{vierbeins}
\end{eqnarray}
\textcolor{black}{Now, we need to determine the $\beta^{\mu}$ matrices as well as the spinorial affine connections $\Omega_{\mu}$ for a spin-1 field in the Eq. (\ref{vbe}). This requires first to construct the space-dependent Dirac matrices. To acquire this,} we can use the following relation: $\gamma^{\mu}=e^{\mu}_{\ a}\overline{\gamma}^{ a}$, in which $\overline{\gamma}^{a}$ are the space-independent Dirac matrices that can be written in terms of the Pauli matrices ($\sigma^{x}, \sigma^{y}, \sigma^{z}$) in $2+1$ dimensions. According to \textcolor{black}{the} signature of the metric in Eq. (\ref{metric}), space-independent Dirac matrices can be chosen as $\overline{\gamma}^{0}=\sigma^{z}$, $\overline{\gamma}^{1}=i\sigma^{x}$, $\overline{\gamma}^{2}=i\sigma^{y}$ and accordingly space-dependent Dirac matrices can be constructed as follows \cite{22}
\begin{eqnarray}
\gamma^{t}=\frac{1}{c}\overline{\gamma}^{0}-\frac{\varpi}{c\alpha r}\overline{\gamma}^{2},\quad \gamma^{r}=\overline{\gamma}^{1},\quad \gamma^{\phi}=\frac{1}{\alpha r}\overline{\gamma}^{2}.\label{matrices}
\end{eqnarray}
\textcolor{black}{To determine the spinorial connections $\Omega_{\mu}$, in the Eq. (\ref{vbe}), we can use the obtained spinorial affine connections for Dirac fields in the Ref. \cite{22},}
\begin{eqnarray}
\Gamma_{t}=0,\quad \Gamma_{r}=0,\quad \Gamma_{\phi}=\frac{i\alpha}{2}\sigma^{z}.\label{spinorialconnections}
\end{eqnarray}
\textcolor{black}{Here, it is clear that there is only one non-vanishing component of the spinorial affine connections and this one is found as $\Omega_{\phi}=i\alpha\ \textrm{diag}\left( 1,0,0,-1\right)$. As a next step, we can introduce the oscillator coupling and this can be acquired through the following non-minimal substitution} \cite{24}
\begin{eqnarray}
&\partial_{r}\rightarrow\partial_{r}+\mathcal{K} \hat{\eta}r,\quad \mathcal{K}= \frac{m\omega_{o}}{\hbar} ,\nonumber\\
& \hat{\eta}= \sigma^{z}\otimes\sigma^{z}=\mathrm{diag}(1,-1,-1,1),
\label{oscillatorcoupling}
\end{eqnarray}
where $\omega_{o}$ is the oscillator frequency (coupling strength) \cite{28}. Also, the presence of an internal magnetic flux can be taken into account through the angular component of the $3$-vector potential, $A_{\phi}=\frac{\phi_{B}}{2\pi\hbar c}$ where $\phi_{B}$ is the internal magnetic flux \cite{37}. Here, we seek the bound state solutions for the system in question and accordingly we assume the interaction is time-independent. \textcolor{black}{This allows us to factorise the spin-1 field as the following}
\begin{eqnarray}
\Psi\left(t,r,\phi\right)=e^{-i\omega t}e^{is\phi}\left(\begin{array}{c}
\psi_{1}\left(r\right)   \\
\psi_{2}\left(r\right)    \\
\psi_{3}\left(r\right)     \\
\psi_{4}\left(r\right)
\end{array}
\right),\label{spinor}
\end{eqnarray}
\textcolor{black}{in which $\omega$ is the relativistic frequency and $s$ is the spin of the test field in question.}

\begin{figure*}
\centering
\subfigure[$\phi_{0}=0.5$, $\varpi=0.5$]{\includegraphics[scale=0.70]{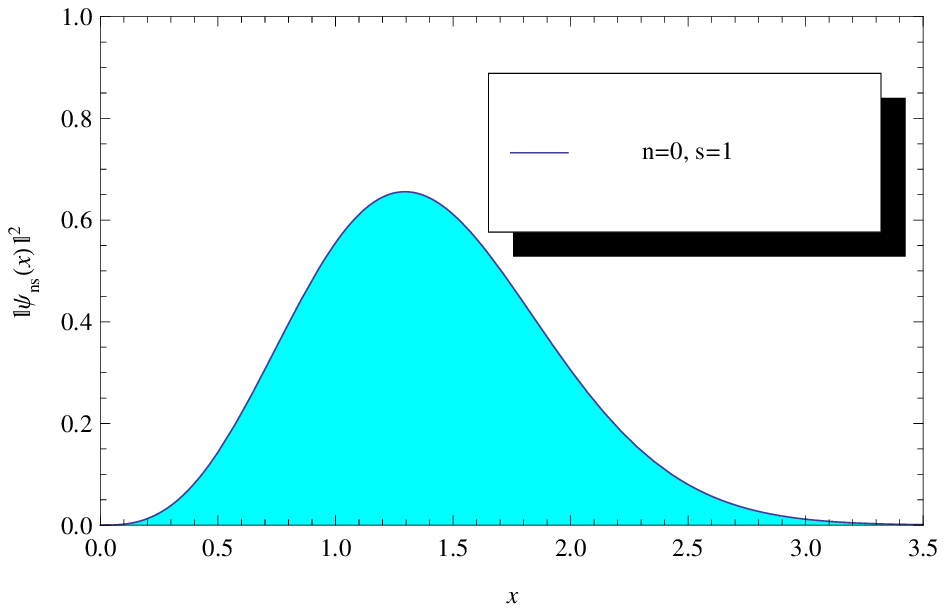}}\quad \subfigure[$\phi_{0}=0.5$, $\varpi=2$]{\includegraphics[scale=0.70]{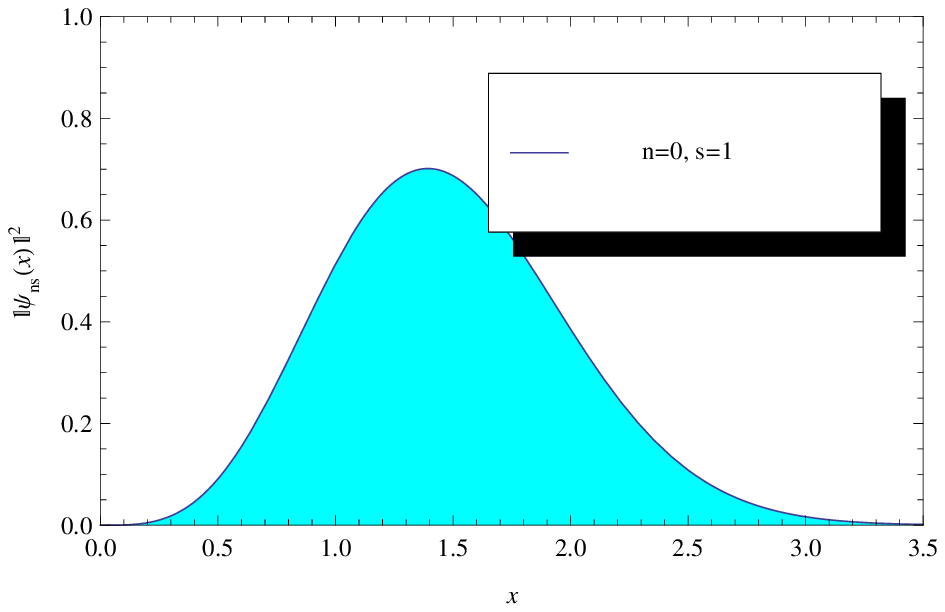}}\\
\subfigure[$\phi_{0}=4$, $\varpi=0.5$]{\includegraphics[scale=0.70]{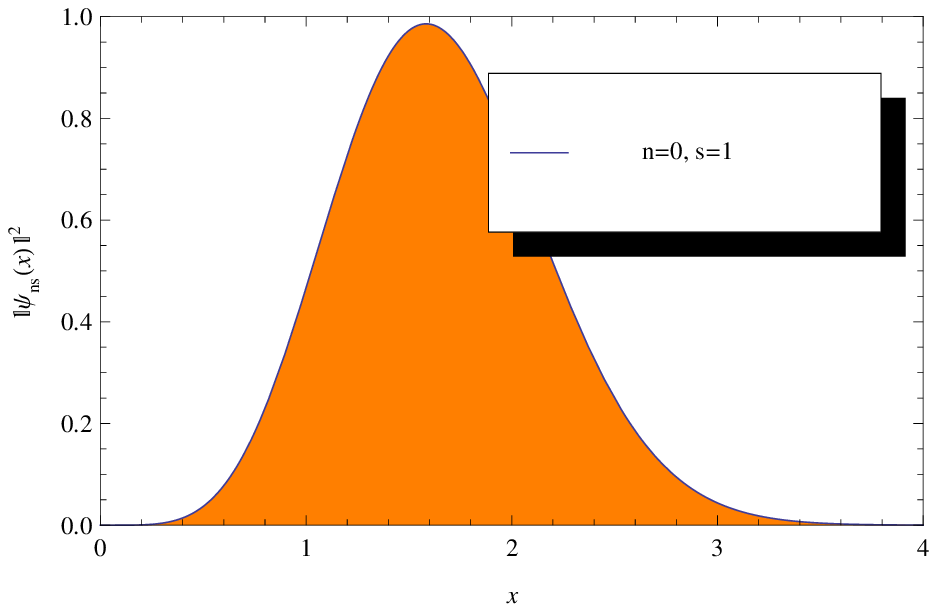}}
\caption{Effects of the internal magnetic flux and spin parameter of the source on the probability density function. Here $e=1$, $c=1$, $\hbar=1$, $m=1$, $\alpha=0.9$, $\omega_{0}=1$, $\pi=1$. }
\label{figb}
\end{figure*}

\section{Altered spectrum}\label{sec3}

In this section, we derive a set of coupled equations for a charged relativistic spin-$1$ oscillator exposed to magnetic flux in the $2+1$ dimensional spacetime generated by a spinning point source and arrive at a second order wave equation in exactly soluble form. By substituting Eq. (\ref{matrices}), Eq. (\ref{spinorialconnections}), Eq. (\ref{oscillatorcoupling}) and Eq. (\ref{spinor}) into the Eq. (\ref{vbe}) we obtain a set of coupled equations which can be rewritten as the following
\begin{eqnarray}
&&\frac{\omega}{c} \psi_{+}\left(r\right)-\tilde{m}\psi_{-}\left(r\right)-\delta(r)\psi_{0}\left(r\right)=0,\nonumber\\
&&\frac{\omega}{c}  \psi_{-}\left(r\right)-\tilde{m}\psi_{+}\left(r\right)-\left(\partial_r - \kappa r\right)\psi_{0}\left(r\right)=0,\nonumber\\
&&\resizebox{7cm}{!}{$\tilde{m} \psi_{0}\left(r\right)+\left(\partial_r +\kappa r +\frac{1}{r}\right)\psi_{+}\left(r\right)-\delta(r)\psi_{-}\left(r\right)=0$},\qquad \label{eqs}
\end{eqnarray}
in which $\delta(r)=\frac{s+\Phi+\tilde{\omega}\varpi}{\alpha r}$,  $\psi_{\pm}\left(r\right)= \psi_{1}\left(r\right)\pm \psi_{4}\left(r\right)$ and $\psi_{0}\left(r\right)= \psi_{2}\left(r\right)+ \psi_{3}\left(r\right)$ since the spinor is a symmetric spinor of rank-two. \textcolor{black}{It can be seen that the first equation in the set of equations, in above, is in fully algebraic form and this makes possible} to obtain easily the following expressions for the components $\psi_{+}$ and $\psi_{-}$
\begin{eqnarray*}
\psi_{+}=-\frac{\tilde{m}\delta(r)}{\tilde{\omega}^2(\frac{\tilde{m}}{\tilde{\omega}}-\frac{\tilde{\omega}}{\tilde{m}})}\psi_{0}-\frac{(\partial_r -\kappa r)}{\tilde{\omega}(\frac{\tilde{m}}{\tilde{\omega}}-\frac{\tilde{\omega}}{\tilde{m}})}\psi_{0}+\frac{\delta(r)}{\tilde{\omega}}\psi_{0},
\end{eqnarray*}
\begin{eqnarray*}
\psi_{-}=-\frac{\delta(r)}{\tilde{\omega}(\frac{\tilde{m}}{\tilde{\omega}}-\frac{\tilde{\omega}}{\tilde{m}})}\psi_{0}-\frac{(\partial_r -\kappa r)}{\tilde{m}(\frac{\tilde{m}}{\tilde{\omega}}-\frac{\tilde{\omega}}{\tilde{m}})}\psi_{0},
\end{eqnarray*}
where $\tilde{\omega}=\frac{\omega}{c}$, $\Phi=\frac{e\phi_{B}}{2\pi\hbar c}$. \textcolor{black}{Accordingly, we can acquire a wave equation $\psi^{\textbf{''}}_{0}+\frac{1}{r}\psi^{\textbf{'}}_{0}+\hat{Q}\psi_{0}=0$, in which
\begin{eqnarray*}
\hat{Q}=[\frac{\tilde{\omega}}{\tilde{m}}\delta^{\textbf{'}}(r)-\delta(r)^2+\frac{2\kappa r^2\tilde{\omega}+\tilde{\omega}}{\tilde{m}r}\delta(r)-\lambda(r)],
\end{eqnarray*}
\begin{eqnarray*}
\lambda(r)=(\kappa^2 r^2+\tilde{m}^2-\tilde{\omega}^2+2\kappa),
\end{eqnarray*}
and prime $^{\textbf{'}}$ denotes derivative with respect to $r$. By defining a dimensionless independent variable, $x=\kappa r^2$, and then considering an ansatz function, $\psi_0\left(x\right)=\frac{1}{\sqrt{x}}\psi\left(x\right)$, to get rid of the term $\propto \psi^{\textbf{'}}\left(x\right)$ one arrives at the following wave equation for the unknown function $\psi\left(x\right)$
\begin{eqnarray}
\psi^{\textbf{''}}\left(x\right)+\left[\frac{q_1}{x}+\frac{\frac{1}{4}-q_{2}^2}{x^2}-\frac{1}{4}\right]\psi\left(x\right)=0,\label{waveq}
\end{eqnarray}
where
\begin{eqnarray*}
q_1 =-\frac{1}{2}+\frac{-\tilde{m}(\tilde{m}^2-\tilde{\omega}^2+2\kappa)\alpha+2\tilde{\omega}\kappa(\tilde{\omega}\varpi+s+\Phi)}{4\tilde{m}\alpha\kappa},
\end{eqnarray*}
and
\begin{eqnarray*}
q_2=\frac{s+\Phi+\tilde{\omega}\varpi}{2\alpha}.
\end{eqnarray*}
Solution function of this wave equation can be expressed in terms of the confluent hypergeometric function $_{1}F_{1}$ as the following
\begin{eqnarray}
\psi\left(x\right)=\mathcal{C}\ x^{\frac{1}{2}+q_{2}} e^{-\frac{x}{2}}\ _{1}F_{1}\left(\tilde{\epsilon}, \tilde{\tau}; x \right),\label{sol.func}
\end{eqnarray}
where $\mathcal{C}$ is an arbitrary constant, $\tilde{\epsilon}=\frac{1}{2}+ q_{2}- q_{1}$ and $\tilde{\tau}=1+ 2q_{2}$. For large values of the argument $x$, the $_{1}F_{1}\left(\tilde{\epsilon}, \tilde{\tau}; x \right)$ function behaves as
\begin{eqnarray*}
_{1}F_{1}\left(\tilde{\epsilon}, \tilde{\tau}; x \right)\approx \frac{\Gamma(\tilde{\tau})}{\Gamma(\tilde{\epsilon})}e^{x}x^{\tilde{\epsilon}-\tilde{\tau}}[1+\mathcal{O}(|x|^{-1}),
\end{eqnarray*}
where $\Gamma$ stands for the Gamma functions \cite{vbo-damp}. That is, it diverges if $x\rightsquigarrow \infty$. Here, our purpose is to determine regular solution to the wave equation and this requires that $\tilde{\epsilon}=-n$ where $n$ is the radial quantum number ($n=0,1,2..$). At that rate, the solution function becomes polynomial of degree $n$ according to $x$ \cite{vbo-damp}. This termination allows us to determine the quantization condition for the considered system and accordingly we obtain an exact result for the relativistic frequency (in closed-form)
\begin{eqnarray*}
\omega_{ns}/c&=&\frac{\kappa}{\tilde{m}\alpha}\ \tilde{\theta}\ \tilde{\eta}^{-1}\nonumber\\
&\pm& \tilde{m}\sqrt{\left[1+\frac{4\kappa}{\tilde{m}^2}n^{\textbf{*}}\right]\tilde{\eta}+\frac{\kappa^2}{\tilde{m}^4\alpha^2}\tilde{\theta}^2 } \ \tilde{\eta}^{-1}\nonumber\\
& \label{spectrum}
\end{eqnarray*}
where $\tilde{\eta}=\left[ 1+\frac{2\kappa\varpi}{\tilde{m}\alpha}\right]$, $\tilde{\theta}=\left[\varpi\tilde{m}-\tilde{\mathcal{S}} \right]$, $\tilde{\mathcal{S}}=s+\Phi$ and $n^{\textbf{*}}=n+1+ \frac{\tilde{\mathcal{S}}}{2\alpha}$. Here, it can be seen that $\tilde{\eta}$ and $\tilde{\theta}$ are dimensionless quantities. Also, writing an explicit expression for the relativistic energy of the considered system may be also useful to compare the findings with the previously announced results. The obtained energy spectra can be written as the following form
\begin{eqnarray}
\mathcal{E}_{ns}&=&\omega_{o}\hbar \frac{\tilde{\theta}}{\alpha} \ \tilde{\eta}^{-1}\nonumber\\
&\pm & mc^2\sqrt{\left[1+\frac{4\omega_{o}\hbar}{m c^2}n^{\textbf{*}}\right]\tilde{\eta}+\frac{\omega_{o}^2\hbar^2}{\tilde{m}^2c^4}\frac{\tilde{\theta}^2}{\alpha^2}} \ \tilde{\eta}^{-1}\nonumber\\
&\label{spectrum-e}
\end{eqnarray}
It is clear that this energy spectra becomes $\mathcal{E}_{n}=\pm mc^2\left\lbrace 1+\frac{4\omega_0\hbar}{mc^2}\left( n+1\right)\right\rbrace^{\frac{1}{2}}$ if $s=0$, $\phi_{B}=0$ and $\varpi=0$. In this case, it can be seen that the considered system never stop oscillating even when it reaches the ground state ($n=0$) and the relativistic energy of such a massive vector field becomes $\mathcal{E}\sim \pm mc^2$ if $\omega_{o}=0$. In $s=0$, $\phi_{B}=0$ and $\varpi=0$ case, we observe that the obtained spectrum  are same (except for mass factor) as previously announced result for the singlet quantum state of a fermion-antifermion pair holding together through $\mathcal{DO}$ interaction \cite{10} and it is also same as the previously announced result for a one-dimensional $\mathcal{DO}$ \cite{pacheco} (see also \cite{41}). However, the energy spectra in the Eq. (\ref{spectrum-e}) is quite different from the obtained non-perturbative result for a $\mathcal{DO}$ system in the spinning cosmic string spacetime \cite{18} (see also \cite{cunha}) when $s\neq0$ and $\varpi\neq0$. Also, as is expected, our results show that the behaviour of a relativistic spin-1 oscillator cannot be same as the behaviour of a relativistic scalar oscillator in the cosmic string spacetime (see Eq. (\ref{spectrum-e}) and \cite{santos}). The non-perturbative spectrum in the Eq. (\ref{spectrum-e}) allows us to discuss the effect of the internal magnetic flux and spacetime parameters besides the oscillator frequency on the relativistic spin-1 field. Now, we can discuss the results in detail. Let we start by considering the $\varpi=0$ and $\Phi=0$ case. Even in this case, we can observe that spin of the considered vector field couples with the oscillator frequency and this coupling is also altered by the string tension. In addition, the effect of the string tension on each possible spin quantum state cannot be same since there exist $s/\alpha$ terms in the spectrum. If we consider only $\Phi=$ case, one can see that the spin of the point source couples not only with mass of the vector boson but also couples with the oscillator frequency. Furthermore, this coupling breaks the symmetry of the energy levels (around the $\mathcal{E}=0$) corresponding to the particle-antiparticle states. On the other hand, it is clear that the magnitude of the relativistic energy increases as the oscillator frequency increases. This makes possible to use such an obtained exact result, for a relativistic spin-1 oscillator, to estimate mass spectra of massive vector bosons, in principle (see also \cite{10}). If there exists magnetic flux acting on such a spin-1 field, it is clear that the spin ($s$) of the vector field is altered by the magnetic flux $\Phi$ in such a way that $s+\Phi$. This seems simple but the presence of the magnetic flux can be responsible for the behaviour of such a particle. In principle, we can realize that the considered system behaves as a fermion if $\Phi$ equals to a half-integer. We can also conclude that the considered vector field may behave as a scalar boson or a vector boson if $\Phi$ equals to an integer. This implies also that statistical behaviours and properties of such relativistic fields can be actively tuned by adjusting the magnetic flux, in principle.}

\section{Summary and discussions}\label{Conc}

In this contribution, we have \textcolor{black}{investigated} the effects of an internal magnetic flux and spin of a point source on a charged relativistic spin-$1$ oscillator in a 2+1-dimensional background generated by a \textcolor{black}{spinning} point source. \textcolor{black}{To do this}, we have solved the associated spin-1 equation for a charged spin-$1$ field with a non-minimal coupling under the influence of an internal magnetic flux in the mentioned spacetime background and have acquired \textcolor{black}{an exact energy spectrum in closed-form. The obtained spectrum is given by the Eq. (\ref{spectrum-e}). This energy expression shows that the topology of the background geometry effects differently the possible spin quantum states ($s=0,\pm 1$) of the system since there exist $s/\alpha$ terms in the obtained spectra. We have also observe that the spin of the considered vector test field} is altered by the internal magnetic flux ($\propto \phi_{B}$) as $s+\frac{e\phi_{B}}{2\pi\hbar c}$ according to the elementary electrical charge ($e$) of the considered relativistic particle. Furthermore, our results show that the altered spin ($\tilde{\mathcal{S}}$) of the considered vector field couples with the \textcolor{black}{oscillator frequency} ($\omega_{0}$) and spin parameter ($\varpi$) of the point source. Here, it is also clear that the info about the string tension is carried by the spin of the spin-1 vector boson and hence the effects of the background on the energy levels of such a test field do not appear if one looks $s=0$ states provided that $\varpi=0$. \textcolor{black}{Even though the internal magnetic flux affects the spin of the considered vector field, this coupling does not mix the positive and negative energy states. We can also see that the oscillator frequency is changed by the altered spin} since the resulting spectrum includes $\tilde{\mathcal{S}}\omega_{0}$ terms. This means that the resulting oscillator may oscillate with a different frequency if there exists an internal magnetic flux acting on the relativistic spin-1 oscillator \textcolor{black}{even if the spacetime is globally and locally flat ($\alpha=1$, $\varpi=0$). In the Figure \ref{figa}, the alterations on the relativistic energy levels can be seen for varying values of the $\phi_{B}$ and $\varpi$ and this figure shows that the magnitude of the relativistic energy levels is changed by $\phi_{B}$ besides the $\varpi$. Here, one can also observe that the $\varpi$ break the the symmetry of the energy levels around the zero energy. In the parts of the Figure \ref{figb}, we have shown the space dependence of the probability density function(s) for different $\phi_{B}$ and $\varpi$ values and these parts show that the $\phi_{B}$ and $\varpi$ parameters can change the magnitude of the probability density function(s). This also implies that the presence of an internal magnetic flux and spin of the point source can impact on particle production (see also \cite{pairproduction2022})}.

\section*{\small{Acknowledgements}}
{\color{black}{The authors thank the kind reviewers for valuable comments and constructive suggestions.}}

\section*{\small{Data availability}}
This manuscript has no associated data or the data will not be deposited.

\section*{\small{Conflicts of interest statement}}

There is no conflict of interest declared by the authors.

\section*{\small{Funding}}

There is no funding regarding this article

\end{document}